\documentstyle[11pt,psfig]{article}

\def\Journal#1#2#3#4{{#1} {\bf #2}, #3 (#4)}

\def\NPB{{\em Nucl. Phys.} B}
\def\PLB{{\em Phys. Lett.}  B}

\def\PRD{{\em Phys. Rev.} D}
\def\RPP{\em Rep. Prog. Phys.}

\def\ds{\displaystyle}

\def\be{\begin{equation}}
\def\ee{\end{equation}}
\def\bea{\begin{eqnarray}}
\def\eea{\end{eqnarray}}

\def\cw{\cos\theta_W}

\def\half{\mbox{\small $\frac{1}{2}$}}
\def \lta {\mathrel{\vcenter
     {\hbox{$<$}\nointerlineskip\hbox{$\sim$}}}}
\def \gta {\mathrel{\vcenter
     {\hbox{$>$}\nointerlineskip\hbox{$\sim$}}}}

\begin{document}
\footnotesep=14pt
\begin{flushright}
\baselineskip=14pt
{\normalsize DFPD-00/TH/11}
\end{flushright}

\vspace*{.5cm}
\renewcommand{\thefootnote}{\fnsymbol{footnote}}
\setcounter{footnote}{0}

\begin{center}
{\large\bf PRIMORDIAL MAGNETIC FIELDS AND \\ELECTROWEAK BARYOGENESIS
\footnote{Talk presented at the Third International Conference on Particle Physics and 
the Early Universe (COSMO-99), Trieste, Italy, 27 Sept - 3 Oct, 1999, to
appear in the proceedings.}
}
\end{center}
\setcounter{footnote}{2}
\begin{center}
\baselineskip=16pt
{\bf Dario Grasso}\footnote{E-mail:
{\tt dario.grasso@pd.infn.it}}\\
\vspace{0.4cm}
{\em Dipartimento di Fisica ``G.Galilei'',
Via F. Marzolo 8, I-35131 Padova, Italy,~ and INFN, Sezione di
Padova}
\end{center}
\baselineskip=20pt
\vspace*{.5cm}
\begin{quote}
\begin{center}
{\bf\large Abstract}
\end{center}
\vspace{0.2cm}
{\baselineskip=10pt%
In this contribution we will shortly review the main
mechanism through which primordial magnetic fields may affect the
electroweak baryogenesis. It is shown that although strong
magnetic fields might enhance the strength of the electroweak
phase transition, no benefit is found for baryogenesis once the
effect of the field on the sphaleron rate is taken into account.
The possible role of hypermagnetic helicity for the electroweak
baryogenesis is shortly discussed. }
\end{quote}
\renewcommand{\thefootnote}{\arabic{footnote}}
\setcounter{footnote}{0}
\newpage
\baselineskip=16pt
\section{Introduction}\label{subsec:1}
An outstanding problem in astrophysics concerns the origin and
nature of magnetic fields in the galaxies and in the clusters of
galaxies \cite{Zeldovich,Kronberg}. The uniformity of the magnetic
fields strength in the several galaxies and the recent observation
of magnetic fields with the same intensity in high red-shift
protogalactic clouds suggest that galactic and
intergalactic magnetic fields may have a primordial origin.
Hopefully, a confirmation  to this intriguing hypothesis will come
from the forthcoming balloon and satellite missions looking at the
anisotropies of the Cosmic Microwave Background Radiation. In
fact, among other very important cosmological parameters, the
observations performed by these surveyors may be able to detect
the imprint of primordial magnetic fields on the temperature and
polarizations acoustic peaks \cite{AdamsDGR}.

Besides observations, a considerable amount of theoretical work,
based on the particle physics standard model as well as on its
extensions, has been done which support the hypothesis that the
production of magnetic fields may actually be occurred during the
big-bang \cite{Enqvist,Ola}.

Quite independently from the astrophysical considerations, several
authors paid some effort to investigate the possible effects that
cosmic magnetic fields may have for some relevant physical
processes which occurred in the early Universe like the big-bang
nucleosynthesis \cite{GraRub96} and the electroweak baryogenesis (EWB).
The latter is the main subject of this contribution.

Since, before the electroweak phase transition (EWPT) to fix the
unitary gauge is a meaningless operation, the electromagnetic
field is undefined above the weak scale and we can only speaks in
terms of hyperelectric and hypermagnetic fields. The importance of
a possible primordial hypercharge magnetic fields for the
electroweak baryogenesis  scenario is three-fold. In fact, we will
show that hypercharge magnetic fields can affect the dynamics of
the EWPT, they change the rate of the baryon number violating
anomalous processes and, finally, hypermagnetic fields may
themselves carry baryon number if they possess a non trivial
topology. These effects will be shortly reviewed  respectively in the section
2,3, and 4 of this contribution.

\section{The effect of a strong hypermagnetic fields on the EWPT}

As it is well known, the properties of the EWPT are determined by
the Higgs field effective potential. In the framework of the
minimal standard model (MSM), by accounting for radiative
corrections from all the known particles and for finite temperature effects,
one obtains that
\be\label{veff}
  V_{\rm eff}(\phi , T) \simeq
-\frac{1}{2}(\mu^2 - \alpha T^2)\phi^2
  -  T \delta \phi^3 + \frac{1}{4}(\lambda - \delta\lambda_T )\phi^4~.
\ee
where $\phi$ is the radial component of the Higgs field and $T$ is the
temperature (for the definitions of the coefficients see {\it e.g.}
Ref.\cite{ElmEK}).

A strong hypermagnetic field can produce corrections to the effective
potential as it affects the charge particles propagators.
It was shown, however, that these kind of corrections are not the most
relevant effect that strong hypermagnetic fields may produce on
the EWPT. In fact, it was recently showed by Giovannini and
Shaposhnikov \cite{GioSha} and by Elmfors, Enqvist and Kainulainen
\cite{ElmEK} that hypermagnetic fields can affect directly the
Gibbs free  energy (in practice the pressure) difference between
the broken and the unbroken phase, hence the strength of the
transition. The effect can be understood in analogy with the
Meissner effect, {\it i.e.} the expulsion of the magnetic field
from superconductors as consequence of photon getting an effective
mass inside the specimen. In our case, it is the $Z$--component of
the hypercharge $U(1)_Y$ magnetic field which is expelled from the
broken phase just because $Z$--bosons are massive in that phase.
Such a process has a cost in terms of free energy. Since in the
broken phase the hypercharge field decompose into
\begin{equation}\label{Ay}
  A^Y_\mu = \cos\theta_w A_\mu - \sin\theta_w Z_\mu~,
\end{equation}
we see that the Gibbs free energy in the broken and unbroken
phases are
\bea
\label{GbGu}
        G_b&=&V(\phi)-\frac 1 2 \cos^2\theta_w (B^{ext}_Y)^2~,\\
        G_u&=&V(0)-\frac 1 2 (B^{ext}_Y)^2~.
\eea
where $B^{ext}_Y$ is the external hypermagnetic field.
 In other words, compared to the case in which no magnetic field is
present, the energy barrier between unbroken and broken phase,
hence the strength of the transition, is enhanced by the quantity
$\displaystyle \frac{1}{2}\sin^2\theta_w (B^{ext}_{Y})^2$.
According to the authors of refs.\cite{GioSha,ElmEK} this effect
can have important consequence for the electroweak baryogenesis
scenario.

In any scenario of baryogenesis it is crucial to know at which epoch do the
sphaleronic transitions, which violate the sum ($B + L$) of the baryon and
lepton numbers, fall out of thermal equilibrium. Generally this happens
at temperatures below $\bar{T}$ such that \cite{Washout}
\be
\frac{E(\bar{T})}{\bar{T}} \ge A\,\,,
\label{washout}
\ee
where $E(T)$ is the sphaleron energy at the temperature $T$ and
$ A \simeq 35 - 45 $, depending on the poorly known prefactor of the
sphaleron rate.
In the case of baryogenesis at the electroweak scale one requires the
sphalerons to drop out of thermal equilibrium soon after the electroweak
phase transition. It follows that the requirement $\bar{T}=T_c$, where $T_c$
is the critical temperature, turns eq. (\ref{washout}) into a lower bound
on the higgs vacuum expectation value (VEV),
\be
\frac{v(T_c)}{T_c} \gta 1\,. \label{first} \ee As a result of
intense research in the recent years \cite{EWPT}, it is by now
agreed that the standard model (SM) does not have a phase
transition strong enough as to fulfill eq. (\ref{first}), whereas
there is still some room left in the parameter space of the
minimal supersymmetric standard model (MSSM).

The interesting observation made in Refs.\cite{GioSha,ElmEK} is that
a  magnetic field for the hypercharge $U(1)_Y$ present for $T>T_c$
may help to fulfill Eq.(\ref{first}). In fact,  it follows from the
Eqs.(\ref{GbGu}), that in presence of the magnetic field the critical
temperature is defined by the expression
\be
\label{Tcrit}
        V(0,T_c)- V(\phi,T_c) =
         \frac{1}{2}\sin^2\theta_w (B^{ext}_{Y}(T_c))^2~.
\ee This expression implies a smaller value of $T_c$ than that it
would take in the absence of the magnetic field, hence a larger
value of the ratio (\ref{first}). On the basis of the previous
considerations and several numerical computations, the authors of
Refs.\cite{GioSha,ElmEK} concluded that for some reasonable values
of the magnetic field strength the EW baryogenesis can be revived
even in the standard model. In the next section we shall reconsider
critically this conclusion.

\section{The sphaleron in a magnetic field}\label{sec:2}
The sphaleron \cite{KM}, is a static and unstable solution of the
field equations of the electroweak model, corresponding to the top
of the energy barrier between two topologically distinct vacua. In
the limit of vanishing Weinberg angle, $\theta_w \to 0$, the
sphaleron is a spherically symmetric, hedgehog-like configuration
of $SU(2)$ gauge and Higgs fields.  No magnetic moment is present
in this case. As $\theta_w$ is turned on the $U_Y(1)$ field is
excited and the spherical symmetry is reduced to an axial symmetry
\cite{KM}. A very good approximation to the exact solution is
obtained using the Ansatz by Klinkhamer and Laterveer \cite{KM},
which requires four scalar functions of $r$ only,
 \bea &&\ds
g^\prime a_i \, dx^i = (1 - f_0(\xi))\, F_3 \,,\nonumber\\ &&\ds g
W^a_i \sigma^a \, dx^i = (1-f(\xi)) (F_1 \sigma^1 +F_2 \sigma^2) +
(1-f_3(\xi)) F_3 \sigma^3 \,,\nonumber\\ &&{\bf \Phi} =
\frac{v}{\sqrt{2}} \left(
\begin{array}{c}
\ds 0 \\
\ds h(\xi)
\end{array}
\right)\,\,, \label{ansatz}
 \eea
 where $g$ and $g^\prime$ are the
$SU(2)_L$ and $U(1)_Y$ gauge couplings, $v$ is the higgs VEV such
that $M_W= g v/2$, $M_h = \sqrt{2 \lambda} v$, $\xi=gvr$,
$\sigma^a$ ($a= 1,2,3$) are the Pauli matrices, and the $F_a$'s
are 1-forms defined in Ref. \cite{KM}. The boundary conditions for
the four scalar functions are \bea
 f(\xi)\,,\,f_3(\xi)\,,\,h(\xi)
\to 0\ \ \ \ f_0(\xi)\to 1\;\; \qquad &&{\mathrm for}\,\;\, \xi\to
0 \nonumber \\
 f(\xi)\,,\,f_3(\xi)\,,\,h(\xi)\,,\,f_0(\xi) \to 1
\qquad &&{\mathrm for}\,\;\, \xi\to \infty\,.\label{bcinf}
\eea
 It is known \cite{KM} that for $\theta_w \neq 0$ the sphaleron
has some interesting electromagnetic properties. In fact,
differently from the pure $SU(2)$ case, in the physical case a
nonvanishing hypercharge current $J_i$ comes-in. At the first
order in $\theta_w$, $J_i$ takes the form
\be
J_i^{(1)} = -  \half g^\prime v^2 \frac{h^2(\xi) [1 -
f(\xi)]}{r^2}\, \epsilon_{3ij} x_j \;, \label{cur_0} \ee where $h$
and $f$ are the solutions in the $\theta_w \to 0$ limit, giving
for the dipole moment
\be
\mu^{(1)} = \frac{2 \pi}{3}  \frac{g^\prime}{g^3 v} \int_0^\infty d\xi
\xi^2 h^2(\xi) [1-f(\xi)]\;.
\label{mu1}
\ee
The reader should note that the dipole moment is a true electromagnetic one
 because in the broken phase only the electromagnetic component of
the hypercharge field survives at long distances.

Following Ref.\cite{noi} we will now consider what happens to the
sphaleron when an external hypercharge field, $A^Y_i$, is turned
on. Not surprisingly, the energy functional is modified as
\be
E = E_0 - E_{\mathrm dip}\,,
\label{energy}
\ee
with
\be
\ds  E_0= \int d^3x \left[ \frac{1}{4} F^a_{ij}F^a_{ij} +
\frac{1}{4} f_{ij}f_{ij} + (D_i {\bf \Phi})^\dagger (D_i {\bf \Phi}) +
V({\bf \Phi}) \right]
\ee
and
\be
\label{intenergy}
E_{\mathrm dip} = \int d^3x J_i A^Y_i = \half \int d^3x f_{ij}f^c_{ij}
\ee
with $f_{ij} \equiv \partial_i A^Y_j -  \partial_j A^Y_i$.
We will  consider here a constant external hypermagnetic field $B^c_Y$
directed along the $x_3$ axis.
In the $\theta_w\to0$ limit the sphaleron has no hypercharge contribution and
then $E_{\mathrm dip}^{(0)} = 0 $.
At $O(\theta_w)$, using (\ref{cur_0}) and (\ref{mu1}) we get a
simple dipole interaction
\be
E_{\mathrm dip}^{(1)} = \mu^{(1)} B^c_Y \; .
\label{lino}
\ee
In order to assess the range of validity of the approximation
(\ref{lino}) one needs to go beyond the leading order in
$\theta_w$ and look for a nonlinear $B^c_Y$--dependence of $E$.
This requires to solve the full set of  equations of motion for
the gauge fields and the Higgs in the presence of the external
magnetic field. Fortunately, a uniform  $B^c_Y$ does not spoil the
axial symmetry of the problem. Furthermore, the equation of motion
are left unchanged ($\partial_i f^c_{ij} =0$) with respect to the
free field case. The only modification induced by $B^c_Y$ resides
in the boundary conditions since -- as $\xi \to \infty$ -- we now
have
\be
 f(\xi)\,,h(\xi) \to 1\,,\;\;\;\;\;\;\;\; f_3(\xi)\,,f_0(\xi) \to 1-B^c_Y
\sin 2\theta_w \frac{\xi^2}{8 g v^2}
\ee
wheras the boundary condition for $\xi \to 0$ are left unchanged.

The solution of the sphaleron equation of motions with the
boundary conditions in the above were determined numerically by
the authors of Ref.\cite{noi}. They showed that  in the considered
$B^c_Y$--range the corrections to the linear approximation
\[
\Delta E \simeq \mu^{(1)} \cw B^c_Y
\]
are less than $5 \%$.
For larger values of $B^c_Y$ non-linear effects increase sharply. However,
for such large magnetic fields the broken phase of the SM is believed to become
unstable to the formation either of $W$-condensates \cite{AO} or of
 a mixed phase \cite{Laine}. In such situations
the sphaleron solution  does not exist any more. Therefore, we
will limit our considerations to safe values $B^c_Y \lta 0.4~T^2$.

From the previous considerations it follows that the conclusion that the
sphaleron freeze-out condition (\ref{washout}) is
satisfied and the baryon asymmetry preserved was premature. Indeed,
in an external magnetic field the relation between the VEV and the sphaleron
energy is altered and Eq. (\ref{first}) does not imply (\ref{washout})
any more. We can understand it by considering the linear approximation
to $E$,
\be
E\simeq E(B^c_Y=0) - \mu^{(1)} B^c_Y \cw \equiv \frac{4 \pi v}{g}
\left(\varepsilon_0 - \frac{\sin 2\theta_w}{g} \frac{B^c_Y}{v^2} m^{(1)}
\right)
\label{newwash}
\ee
where $m^{(1)}$ is the $O(\theta_W)$ dipole moment expressed in
units of $e/\alpha_W M_W(T)$.
From the Fig.1 we see that even if $v(T_c)/T_c \gta 1$ the washout condition
$E/T_c \gta 35$ is far from being fulfilled.
\begin{figure}[b]
\vspace{9pt}
\centerline{\hbox{\psfig{figure=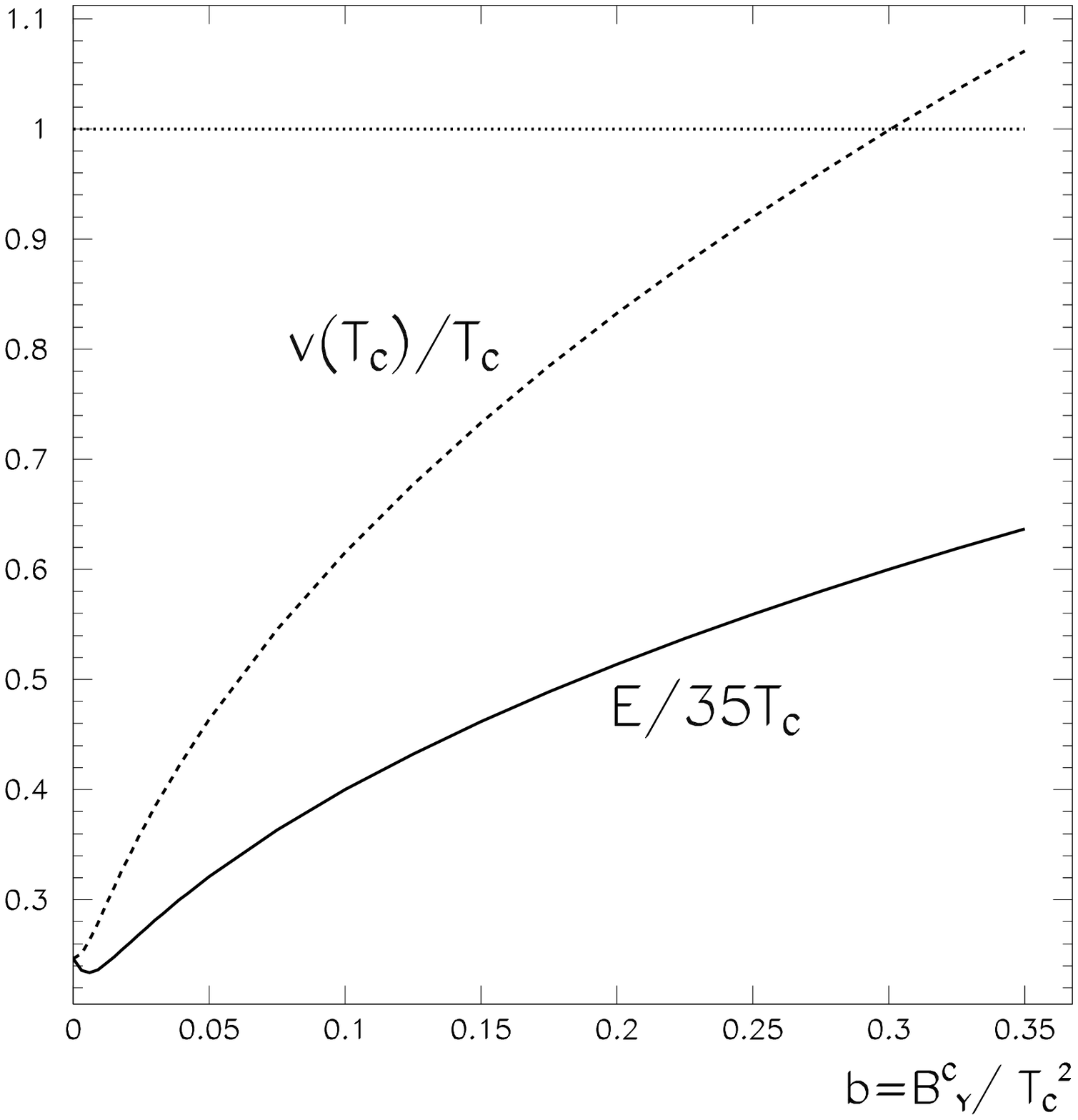,width=7cm}}}
\caption{The VEV at the critical temperature, $v(T_c)$, and the
sphaleron energy {\em vs.} the external magnetic field for
$M_h=M_W$.}
\end{figure}
It follows that the presence of a strong {\it homogeneous}
hypermagnetic field does not seem to help the EWB.

\section{Baryons from hypermagnetic helicity}

A more interesting scenario may arise if hypermagnetic fields are
inhomogeneous and carry a nontrivial topology. The topological
properties of the hypermagnetic field are quantified by the, so
called, hypermagnetic helicity, which coincide with the
Chern-Simon number
\be
N_{CS} = \frac \alpha \pi \int_V d^3x B_Y \cdot A_Y
\ee
 where $A_Y$ is the hypercharge field. It is well known that the
Chern-Simon number is related to the lepton and baryon number by
the abelian anomaly. Recently it was noticed by Giovannini and
Shaposhnikov \cite{GioSha} that the magnetic helicity may have
some non-trivial dynamics during the big-bang giving rise, through
the anomaly equation, to a variation of the fermion and baryon
contents of the Universe.
 Magnetic configurations with $B_Y\cdot {\vec \nabla}\times B_Y$ (
``magnetic knots'' ) may have been produced in the early Universe,
for example, by the conformal invariance breaking coupling of a
pseudoscalar field with the electromagnetic field which may arise,
for example, in some superstring inspired models \cite{GioSha,Veneziano}.

The interesting point that we would like to arise in the
conclusions of this contribution is that a less exotic source of
hypermagnetic helicity is provided by the Weiberg-Salam model
itself. This source are electroweak strings. Electroweak strings
are well known non-perturbative solutions of the Weinberg-Salam
model (for a review see \cite{AchVac}). They generally carry a
nonvanishing Chern-Simon number and, according to recent lattice
simulations, they are copiously produced during the EWPT even if
this transition is just a cross-over \cite{Chernodub}. Although
electroweak string have been sometimes invoked for alternative
mechanism of EWB, it was not always noticed in the literature that
CP symmetry is naturally broken for twisted electroweak strings
without calling for extension of the Higgs structure of the model.
This is just because the twist give rise to non-orthogonal
hyperelectric and hypermagnetic fields. We suggest that primordial
magnetic fields, even if they are uniform, could provide a bias
for the baryon number violation direction. Finally, electroweak
string decay might provide the third Sacharov ingredient for EWB,
namely an out-of-equilibrium condition.

In conclusions, we think that a more careful study of the possible
role of electroweak strings for the EWB is worthwhile.


\end{document}